\documentclass[twocolumn,superscriptaddress,prl,aps,floats,showpacs]{revtex4-1}
\usepackage{amsmath,amssymb,amsfonts,amsthm}
\usepackage{graphicx}
\usepackage{natbib}
\usepackage{bm}
\usepackage{color}

\newcommand{\Cov}{\mathrm{Cov}}

\begin{document}
\title{Schramm-Loewner Evolution and Liouville Quantum Gravity}
\author{Bertrand Duplantier}
\affiliation{Institut de Physique Th\'{e}orique, CEA/Saclay, F-91191
Gif-sur-Yvette Cedex, France}

\author
{Scott Sheffield}
\affiliation{Department of Mathematics,  Massachusetts Institute of Technology,
Cambridge, MA 02139, USA}

\date{\today}

\begin{abstract}

We conformally weld  (via ``quantum zipping'') two boundary arcs of a
Liouville quantum gravity random surface to generate a random curve called the Schramm-Loewner evolution (SLE).  We
develop a theory of \textit{quantum fractal measures} (consistent with the
Knizhnik-Polyakov-Zamolochikov relation) and analyze their evolution under welding
 via SLE martingales.  As an application, we
construct the natural quantum length and boundary intersection measures on
the SLE curve itself.
\end{abstract}
\maketitle

\textit{Introduction.}---Thirty years ago, Polyakov \cite{MR623209}  invented the now celebrated model of \textit{Liouville 2D quantum gravity}, giving the first mathematical description of \textit{continuous random surfaces}, and of the summation over random Riemannian metrics involved.
 While the alternative \textit{discrete} representation by random planar graphs was developed via random matrix theory, the convergence of its continuous limit to Liouville quantum gravity (still not yet rigorously proven) became clear only after Knizhnik, Polyakov and Zamolodchikov (KPZ) \cite{MR947880,MR981529,*MR1005268} proposed their famous relation between critical exponents on a random surface and in the Euclidean plane. Via KPZ, Kazakov's exact solution of the Ising model on a random planar graph \cite{1986PhLA..119..140K}   indeed matched Onsager's results in the plane. The KPZ relation was recently rigorously proven \cite{2008arXiv0808.1560D,*2009arXiv0901.0277D}.

The theory of critical phenomena in the plane, on the other hand, is well-known to be related to \textit{conformal field theory} (CFT)  \cite{MR757857}, a discovery somehow anticipated in the so-called Coulomb gas approach to critical  2D statistical models (see, e.g., \cite{1984JSP....34..731N}). Introduced in 1999,  \textit{Schramm-Loewner evolution} (SLE) \cite{MR1776084}  provided a direct mathematical construction of the \textit{universal continuous scaling limit} of 2D critical curves, an outstanding  invention historically on par with Wiener's  1923 first mathematical construction of universal continuous Brownian motion.

While Liouville field theory, itself a CFT, can be heuristically coupled to other CFT's via KPZ and the so-called \textit{conformal Ansatz} \cite{MR947880,MR981529,*MR1005268,Ginsparg-Moore}, and quantum gravity used to predict properties of critical curves and SLE \cite{2000PhRvL..84.1363D,*MR1964687}, and while SLE can be related to standard CFT \cite{2003CMaPh.239..493B,*2003CMaPh.243..105F}, %,*MR2112128
there is as yet no direct rigorous relationship between Liouville quantum gravity and Schramm-Loewner evolution. We  present such an explicit relationship here, using a conformal welding related to a conjecture by P. Jones (see \cite{sheffield-2010} for details). We give a natural quantum gravity interpretation of certain related SLE \textit{martingales}. The Liouville-SLE relationship so obtained is the rigorous analog of the conformal Ansatz for central charge in the Liouville-CFT correspondence \cite{MR947880,MR981529,*MR1005268,Ginsparg-Moore}.
    We also construct  \textit{quantum gravity fractal measures} for SLE using the KPZ formula. (See \cite{2007arXiv0709.3664K,*BW,AJKS} for some related ideas.)

\textit{Liouville quantum gravity.}---(Critical) Liouville quantum gravity consists of changing the (Lebesgue) area measure $dz$ in a domain $\mathcal D \subset \mathbb C$ to the \textit{quantum area measure}
 $d\mu_\gamma(z):=e^{\gamma h(z)} dz$, where $\gamma$ is a real parameter and where $h$ is an instance of the
(zero boundary, for now) Gaussian free field (GFF),  with
 Dirichlet energy
$(h,h)_{\nabla}:=({2\pi})^{-1}
\int_{\mathcal D} \nabla h(z) \cdot \nabla h(z)dz.$
For $0\leq \gamma <2$ this allows us to mathematically define a
 quantum random surface ${\mathcal S}:=(\mathcal D,h)$ \cite{2008arXiv0808.1560D,2009arXiv0901.0277D}, even though $h$ is actually a distribution \cite{MR2322706}. The
 {measure} $d\mu_\gamma(z)$
can be constructed as the limit as $\varepsilon \to 0$ of the regularized quantities
$d\mu_{\gamma,\varepsilon}(z):=\varepsilon^{\gamma^2/2}\exp[\gamma h_\varepsilon(z)] dz$, where
$h_\varepsilon(z)$ is the mean value of $h$ on the circle $\partial B_\varepsilon(z)$, boundary of the ball
$B_\varepsilon(z)$ of radius $\varepsilon$ centered at $z$; note in particular
 that	$\mathbb E\,e^{\gamma h_\varepsilon(z)}=\big[C(z;\mathcal D)/\varepsilon\big]^{\gamma^2/2}$\cite{2008arXiv0808.1560D,2009arXiv0901.0277D}, where $C(z;\mathcal D)$ is the conformal radius of $\mathcal D$ viewed from $z$.

\textit{Quantum fractal measures and KPZ.}---Consider  $d$-dimensional \textit{Euclidean} or analogously \textit{quantum measures}  of
 planar \textit{fractal} sets
and their a priori scaling properties:\\
$\bullet$ If we rescale a $d$-dimensional fractal $X \subset \mathcal D$  via
the map $z\to \phi(z)=b z$, $b\in \mathbb C$ (so that the Euclidean area of $\mathcal D$ is multiplied by $|b|^2$) then the
$d$-dimensional Euclidean fractal measure of $X$ is multiplied by $|b|^{d} = |b|^{2-2x}$, where $x$ (the so-called
 \textit{Euclidean scaling weight}) is defined by $d := 2-2x(\leq 2)$.\\
$\bullet$  If $X$ is a fractal subset of a random surface $\mathcal S$, and we rescale  $\mathcal S$
so that its quantum area increases by a factor of $|b|^2$, then the quantum fractal measure of $X$
is multiplied by $|b|^{2-2\Delta}$, where $\Delta$ is the analogous \textit{quantum scaling weight}.

	The above assertions suggest
that the ($\gamma$-dependent) Liouville quantum measure $\mathcal Q(X,h)$ of a fractal $ X \subset \mathcal D$ should satisfy the following scaling axioms:\\
$\bullet$ Adding a constant $\lambda_0$ to $h$ (which has the effect of multiplying the quantum area by
$e^{\gamma \lambda_0}$) should cause the fractal measure to be multiplied by $(e^{\gamma \lambda_0})^{1-\Delta}
= e^{\alpha \lambda_0}$:
\begin{eqnarray} \label{h+lambda0}
\mathcal Q(X, h+\lambda_0) &=& e^{\alpha \lambda_0} \mathcal Q(X, h) \\ \label{alphadelta}
\alpha &:=& \gamma - \gamma \Delta.
\end{eqnarray}
$\bullet$ If $\phi(z) = bz$, then
\begin{eqnarray}\label{Alambda}
 \mathcal Q( \phi(X), h \circ \phi^{-1}) = |b|^{d+\alpha^2/2} \mathcal Q(X, h).
\end{eqnarray}
To explain \eqref{Alambda}, note that if we can cover $X$ by $\mathcal N$ radius-$\varepsilon$ balls, then it will take
$\mathcal N |b|^{d}$ such balls to cover $b X$.
 One next observes that the law of $\underline{h}(\cdot):=h(\cdot) - h_\varepsilon(z)$ on $B_\varepsilon(z)$, given $h_\varepsilon(z)$,
is an appropriately projected GFF on a disc, which is independent of $h_\varepsilon(z)$ and $z$ (up to negligible effects of $\partial \mathcal D$; see \cite{2008arXiv0808.1560D}), so one can apply
\eqref{h+lambda0}  to $\underline{h} +\lambda_0$, with the local shift $\lambda_0=h_\varepsilon(z)$. Then the expected resulting conformal factor
$\mathbb E\, e^{\alpha h_\varepsilon(z)}$  will be
$|b|^{\alpha^2/2}$ times larger in the domain $b\mathcal D$,
 because of  the conformal radius  $C(bz;b \mathcal D)= |b|\, C(z;\mathcal D)$. Thus the expected (w.r.t.\ $h$)  quantum measure of $bX$ within one of the $\varepsilon$-balls
covering $bX$ (near $bz$) should be $|b|^{\alpha^2/2}$ times that of $X$ within one of the $\varepsilon$-balls covering $X$
(near $z$).
The law of large numbers for the covering yields \eqref{Alambda}.\\
$\bullet$ $\mathcal Q( \phi(X), h \circ \phi^{-1} - Q \lambda) = \mathcal Q(X, h)$  for $\phi(z)=bz,\,\lambda =\log |b| = \log|\phi'|$. This is because in general
(see \cite{2008arXiv0808.1560D,Ginsparg-Moore}), the pair $\mathcal S=(\mathcal D,h)$ describes the same quantum surface (up to
\textit{coordinate change}) as the
conformally transformed pair $(\phi (\mathcal D), h \circ \phi^{-1} - Q \log|\phi'|)$, and
\begin{equation}\label{Qgamma0}
Q:={\gamma}/{2}+{2}/{\gamma}.
\end{equation}
These properties taken together imply that
\begin{equation}\label{xalpha}
d = \alpha Q - {\alpha^2}/{2},
 \end{equation}
which by \eqref{alphadelta} and \eqref{Qgamma0} is equivalent to the celebrated KPZ formula  \cite{MR947880}: $x=({\gamma^2}/{4})\Delta^2+\left(1-{\gamma^2}/{4}\right)\Delta$.
  Indeed, the above may be viewed as a rather heuristic but genuine derivation of that formula.

 \begin{figure}
\begin{center}
\includegraphics[angle=0,width=.93290\linewidth]{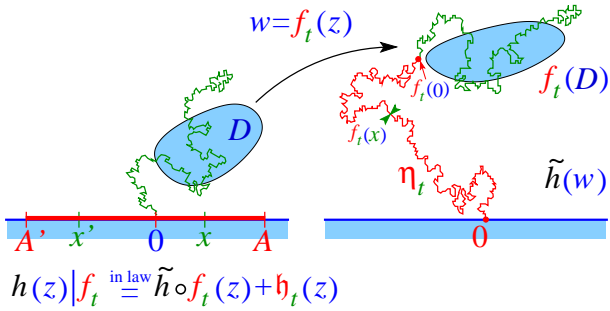}
\end{center}
\vskip-.5cm\caption{\label{slegffD} Reverse chordal $\textrm{SLE}_\kappa$ map $w=f_t(z)$ in half-plane $\mathbb H$.
 Conditioned on  $f_t$,  $h$
is the pullback $\tilde h\circ f_t$ of an unconditioned free boundary GFF $\tilde{h}$, plus the martingale
${\mathfrak h}_t$ \eqref{ht}. For any pair $(x',x)$
of boundary points  mapped to  $f_t(x')=f_t(x)$ on the SLE trace $\eta_t$, the
  \textit{boundary} \textit{quantum lengths} associated with  $h$ of the two conformally welded segments $[x',0]$ and $[0,x]$ are equal \cite{sheffield-2010}. The $\textrm{SLE}_\kappa$  $X=\eta$  on the left is independent of $h$.}
\end{figure}
\textit{SLE definition}.---In its so-called chordal version, the \textit{Schramm-Loewner evolution}
\cite{MR1776084} describes
 the uniformizing conformal map $g_t:\mathbb H\setminus K_t \to \mathbb H$, from the half-plane slit by the external envelope $K_t$ of
 the trace $\eta([0,t])$ up to time $t \geq 0$ of the ${\rm SLE}_\kappa$ path $\eta(t)$ to $\mathbb H$ itself.  This map
satisfies the stochastic differential equation (SDE) $
dg_t(z)=2 dt/g_t(z)-\sqrt \kappa d{B}_t
$, where $B_t$ is standard Brownian motion with $B_0=0$, and  $\kappa \geq 0$. One has $g_t(\eta(t))=0$, while for $t=0$, $g_0(z)=z$.
 For values $0\leq \kappa \leq 4$,  the
$\textrm{SLE}_{\kappa}$ trace is a simple curve (so that $K_t=\eta([0,t])$), while for  $4< \kappa < 8$ it develops double points and becomes space-filling for $\kappa \geq 8$ \cite{MR2153402}. Of particular physical interest are the cases
of the \textit{loop-erased random walk}
($\kappa=2$) \cite{MR2044671}, the  \textit{self-avoiding walk} ($\kappa=8/3$, still conjectural from a rigorous perspective),
the \textit{Ising model interface} ($\kappa=3$ or $16/3$) \cite{MR2680496,*2009arXiv0910.2045C}, the \textit{GFF contour lines} ($\kappa=4$) \cite{MR2486487}, and the \textit{percolation interface} ($\kappa=6$) \cite{MR1851632}.

%For $\kappa \leq 4$,
Let $f_t: \mathbb H\to \mathbb H \setminus K_t$ be the \textit{reverse (``zipping up'')} SLE conformal map (Fig. \ref{slegffD}) given by the SDE $
df_t(z)=-2 dt/f_t(z)-\sqrt \kappa dB_t,
$
with $f_0(z)=z$, $f_0'(z)=1$.  Differentiating w.r.t.\ $z$ gives
$
df'_t(z)=2 dt f'_t(z)/f^2_t(z)$. For each $t\geq 0$, $f_t$ maps $z\in \mathbb H$ to $w:=f_t(z)\in \mathbb H \setminus
\eta_t$ for an SLE$_\kappa$ segment $\eta_t$ with tip $f_t(0)$.

\textit{A (reverse) SLE martingale}.---Define a real stochastic process for $t \geq 0$ and $z \in \mathbb H$, by
\begin{eqnarray}
\label{h0}
{\mathfrak h}_0(z)&:=&({2}/{\sqrt \kappa})\log
|z|\\\label{ht} {\mathfrak h}_t(z)&:=&\mathfrak{h}_0\circ f_t(z)
+Q\log|f'_t(z)|.
\end{eqnarray}
By standard stochastic It\^o calculus,
 this process is a (local) martingale (so that $\mathbb E {\mathfrak h}_t(z)={\mathfrak h}_0(z)$)
for
\begin{equation}
\label{Qkappa} Q={\sqrt \kappa }/{2}+{2}/{\sqrt \kappa},
\end{equation}
for which $
d{\mathfrak h}_t(z)=- R_t(z)d{B}_t$, with $R_t(z):=\Re[{2}/{f_t(z)}]$. The standard quadratic variation of Brownian motion is $\langle dB_t,dB_t\rangle = dt$,
hence $\langle d{\mathfrak h}_t(y),d{\mathfrak h}_t(z)\rangle =R_t(y)R_t(z) dt$. Let us now introduce  the standard \textit{Neumann Green function} in the half-plane $\mathbb H$,
$G_0(y,z):=-\log (|y-z||y-\overline z|),$
 such that $-\Delta G_0(y,z)=2\pi \delta (y-z)$,
with Neumann boundary conditions
${\partial G_0(y,z)}/{\partial \Im y}=0$
for $y\in\mathbb R$.
Define the \textit{time-dependent} Green function $G_t(y,z):=G_0\big(f_t(y),f_t(z)\big)$, i.e.,
  $G_0$ taken at the image points under $f_t$.
A direct calculation shows that the quadratic variation
 above can then simply be written as
the {Green function's} variation (Hadamard's formula)
$ \langle d{\mathfrak h}_t(y),d{\mathfrak h}_t(z)\rangle =-d G_t(y,z).$
Integrating with respect to $t$ yields the covariation of the ${\mathfrak h}_t$ martingale
$\langle {\mathfrak h}_t(y),{\mathfrak h}_t(z)\rangle =G_0(y,z)- G_t(y,z).$
Taking the limit $y\to z$ in the latter, one obtains
\begin{eqnarray}\label{qvar}
\langle {\mathfrak h}_t(z),{\mathfrak h}_t(z)\rangle =C_0(z)-C_t(z),
\end{eqnarray}
where  $C_t(z):=-\log\big[\Im f_t(z)|f'_t(z)|\big]$.

\textit{Quantum conformal welding.}---Consider $h:=\tilde{h} + {\mathfrak h}_0$,
sum of an instance $\tilde{h}$ of the Gaussian free field on $\mathbb H$
with \textit{free boundary conditions} (f.b.c.)\ on $\mathbb R$ (up to additive constant), and of the deterministic function ${\mathfrak h}_0$ \eqref{h0}.
 This $h$ can be coupled \cite{sheffield-2010} with the reverse Loewner
flow evolution $f_t$ described above  so that, given $f_t$, the conditional law of $h$ (hereafter denoted by $h|f_t$) is
\begin{eqnarray}\label{inlaw}
h(z)|f_t\stackrel{\rm in\,\, law}{=}\tilde{h}\circ f_t(z)+{\mathfrak h}_t(z),
\end{eqnarray}
where  $\tilde{h}\circ f_t$ is the pullback of the free boundary GFF $\tilde h$ in the image half-plane, and where ${\mathfrak h}_t$ is the martingale \eqref{ht}. To sample $h$, one can first sample the $B_t$ process (which determines $f_t$), then sample independently the f.b.c.\ GFF $\tilde h$ and take \eqref{inlaw}.
 Its conditional expectation  w.r.t.\ $\tilde h$  is the martingale $\mathbb E\big[h(z)|f_t\big]={\mathfrak h}_t(z)$.
  {Owing to \eqref{ht} the r.h.s.\ of \eqref{inlaw} is of the form
$h\circ f_t+Q\log |f'_t|$. For $Q$ equal to \eqref{Qgamma0}, this is the \textit{transformation law} in Liouville quantum gravity of  the GFF  $h$  under the conformal map $f_t$ \cite{2008arXiv0808.1560D,Ginsparg-Moore}.   Then the pair $(\mathbb H, {\tilde h}\circ f_t+\mathfrak h_t)$ describes the same random surface as the pair $(\mathbb H\setminus K_t,h)$:   Given $f_t$, the image under $f_t$ of the measure $e^{\gamma h(z)}dz$ in $\mathbb H$ is
a random measure whose law is  the \textit{a priori} (unconditioned) law of $e^{\gamma h(w)}dw$ in $\mathbb H\setminus K_t$.} From  \eqref{Qgamma0} and (\ref{Qkappa})
we find  the two dual solutions  $\gamma:=\sqrt{\kappa \wedge \kappa'}$ or $\gamma':=4/\gamma=\sqrt{\kappa\vee \kappa'}$, with $\kappa':=16/\kappa$.
The first solution $\gamma\leq 2$  corresponds precisely to the famous \textit{conformal Ansatz}
\cite{MR947880,MR981529,*MR1005268,Ginsparg-Moore}, that relates the parameter $\gamma =\big(\sqrt{25-c}-\sqrt{1-c}\big)/{\sqrt{6}}$ in Liouville theory to the \textit{central charge} $c=\frac{1}{4}(6-\kappa)(6-\kappa')$ of the CFT coupled to gravity.
The second solution $\gamma'=4/\gamma \geq 2$
corresponds to a \textit{dual} model of Liouville quantum gravity, in which the quantum area measure develops atoms with localized area \cite{2009arXiv0901.0277D,1995PhRvD..51.1836K}.

 The equality in law \eqref{inlaw} essentially results from using in the above $G_0(y,z)=\Cov [\tilde h(y),\tilde h(z)]$ (thus $G_t=\Cov [\tilde h\circ f_t,\tilde h\circ f_t]$), and the fact that ${\mathfrak h}_t(y){\mathfrak h}_t(z)+G_t(y,z)$ is a  martingale \cite{sheffield-2010}.
In this particular coupling of $h$ and $f_t$, the two strands   of the boundary to be matched along the trace $\eta_t$
when ``zipping-up" by the reverse Loewner map $f_t$ have the
 \textit{same quantum length} (at least for $\kappa < 4$) (Fig. \ref{slegffD}).
 This \textit{quantum conformal welding} property actually determines $f_t$ as a function of $h$
 \cite{sheffield-2010}.

 Let $X=\eta$ be an SLE$_\kappa$ independent of $h$ (Fig. \ref{slegffD}).  Define its \textit{``zipping down''} map by $f_{-t}:= g_t: \mathbb H \setminus \eta([0,t]) \to \mathbb H$, $t\geq 0$.  When $\kappa < 4$, $X$ divides $\mathbb H$ into a pair of welded quantum surfaces  that is  \textit{stationary} w.r.t.\ zipping up or down via the transformations $f_t$ ($t \in \mathbb R$) \cite{sheffield-2010}.
\textit{The relation between $\gamma$ and $\kappa$ is now rigorously clear: conformally welding two $\gamma$-quantum surfaces produces}  SLE$_\kappa$.

\textit{Exponential martingales}.--- Let us introduce the {conditional expectations} of exponentials of the field \eqref{inlaw}, ${\mathcal M}^{\alpha}_t(z):=\mathbb E\big(e^{\alpha h(z)}|f_t\big)$, depending on a real parameter $\alpha$,  which are fundamental objects describing quantum gravity coupled to the SLE process.
These \textit{martingales} can be given explicitly  in terms of  \eqref{ht} and \eqref{qvar}:
\begin{eqnarray} \label{expM} \mathcal M^{\alpha}_t(z)&=&\exp\big[\alpha
{\mathfrak h}_t(z)+({\alpha^2}/{2}) C_t(z)\big]\\
\label{expM2}
&=&|w|^{2\alpha/\sqrt\kappa}\left\vert{f'_t}{(z)}\right\vert^{\alpha Q - \alpha^2/2}({\Im w})^{-\alpha^2/2},
\end{eqnarray}
where $w=f_t(z)$. Because of \eqref{qvar},
  \eqref{expM} is an   \textit{exponential martingale with respect to the Brownian motion driving the reverse SLE process:}
\begin{equation}\label{Mzero}
\mathbb E  \mathcal M^{\alpha}_t(z)=\mathcal M^{\alpha}_0(z)=|z|^{2\alpha/\sqrt{\kappa}} (\Im z)^{-\alpha^2/2}.
\end{equation}
A stronger statement is the identity in law of the conditional exponential measure
\begin{eqnarray}\label{inlawexp}
\big(e^{\alpha h(z)}|f_t\big)\, dz \stackrel{\rm in\,\, law}{=}|f'_t(z)|^{d-2} e^{\alpha h(w)} dw,
\end{eqnarray}
with $d$ given by \eqref{xalpha}, $d w=|f'_t(z)|^{2} dz$, and whose expectations \eqref{expM2} agree.

\textit{Expected quantum area.}---For $\alpha=\gamma$, i.e.,  $d=2$ in \eqref{xalpha} and \eqref{inlawexp},
we get   in  \eqref{expM2} the invariant (expected) quantum area  $d\mathcal A:=\mathcal M_t^{\gamma}(z) dz=\mathcal M_0^{\gamma}(w) dw$,
for $\gamma=\sqrt \kappa \wedge 4/\sqrt \kappa$:
\begin{eqnarray}\nonumber
d{\mathcal A}=dz\, \mathbb  E [e^{\gamma h(z)}|f_t]&=&dw\, |w|^{2-\kappa/2} ({\sin \varphi})^{-\kappa/2},\kappa \leq 4\\ \nonumber
&=&dw ({\sin \varphi})^{-8/\kappa},\kappa  \geq 4;\varphi:=\arg w .
\end{eqnarray}
We now  construct explicit invariant SLE quantum measures, using the martingales \eqref{expM} for $\alpha\neq\gamma$.

\textit{SLE bulk quantum measure.}---An SLE measure recently introduced in the context of the so-called  \textit{natural parametrization} of SLE \cite{2009arXiv0906.3804L} describes the  ``fractal length''  of the intersection $X\cap D$ of an (infinite) $\mathrm{SLE}_\kappa$  fractal path  $X=\eta$ with an arbitrary domain $D \subset \mathbb H$ (Fig. \ref{slegffD}-left).  It is shown in  \cite{2009arXiv0906.3804L} that its expectation with respect to the SLE law is finite for any bounded $D$, and given by
$\nu(D):=\int_D  G(z) dz,$
where $G(z):=|z|^{a}|\Im z |^{b}$, with $a=1-8/\kappa$, $b=8/\kappa+\kappa/8-2$. Under a zipping-up conformal map $f_t$,  the expectation $\nu_{f_t} (D_t )$ of the fractal length of the image path $f_t({X})$ in the image domain $D_t:=f_t(D)$ (Fig. \ref{slegffD}-right)
is \textit{conformally covariant}:
\begin{eqnarray} \label{cov}
\nu_{f_t} (D_t )=\int_{D} |f'_t(z)|^d G(z) dz
=\int_{D_t} N_t(w) dw,
\end{eqnarray}
where  $d$ is the $\mathrm{SLE}_\kappa$  \textit{Hausdorff dimension},  equal to $d:=1+\kappa/8$ \cite{MR2435854}, and $N_t(w):=G(z) \vert f'_t(z)\vert^{d-2}$, with $z=f_t^{-1}(w)$.
Replacing $f_t^{-1}$ by the zipping-down map $f_{-t}$, we observe that $M_t:= (G \circ f_{-t}) |f_{-t}'|^{2-d}$ describes the density of expected Euclidean fractal length of $X=\eta$ \textit{given} $\eta([0,t])$ \cite{2009arXiv0906.3804L}.  This $M_t$ is a local martingale w.r.t.\ the forward direction SLE flow $f_{-t}$ that generates $X=\eta$ \cite{2009arXiv0906.3804L}.  Thus $\int_{D\setminus \eta([0,t])} M_t(z)dz$ is a martingale minus the length of $\eta([0,t]) \cap D$; this unique \textit{Doob-Meyer decomposition} actually determines the latter length
 \cite{2009arXiv0906.3804L}.

We  extend this construction to the quantum case by defining
 the expected (w.r.t.\ $X$, given $h$) Liouville \textit{quantum length} $\nu_{\mathcal Q}$ of an infinite SLE path in a domain $D$
\begin{equation}\label{Qtheta}
\nu_{\mathcal Q} (D,h):=\int_D e^{\alpha h(z)}G(z) dz,
\end{equation}
where $\alpha:=\sqrt \kappa/2$ ($=\gamma/2$ for $\kappa \leq 4$, and $\gamma'/2$ for $\kappa > 4$) is chosen to satisfy KPZ  \eqref{xalpha} for the SLE dimension $d=1+\kappa/8$
 (and Seiberg's bound $\alpha \leq Q$ \cite{1990PThPS.102..319S,2009arXiv0901.0277D}).  Conditioning \eqref{Qtheta} on
 $f_t$ and using \eqref{inlawexp} gives
  \begin{eqnarray}  \nonumber \nu_{\mathcal Q}|f_t:=\int_{D} \big(e^{\alpha h(z)}|f_t\big)  G(z) dz\stackrel{\rm in\,\, law}{=}
  \int_{D_t} e^{\alpha h(w)}N_t(w) dw .
 \end{eqnarray}
This formula for the \textit{conditional expected} measure yields, by Doob-Meyer decomposition,  an implicit construction of the quantum length measure. It exists from \cite{2010arXiv1006.4936L,*LWess} since  the second moment $\mathbb E [e^{\alpha h(y)+\alpha h(z)}M_t(y) M_t(z)]$   is bounded by  $|y-z|^{{\mathfrak d}-2}$, with  ${\mathfrak d}:=d-\alpha^2=1-\kappa/8$, thus integrable for ${\mathfrak d}>0$, i.e., $\kappa<8$;
 this measure coincides with the one defined on $\mathbb R$ by unzipping via  $f_{-t}$ \cite{2008arXiv0808.1560D,sheffield-2010} under a finite expectation assumption \cite{2009arXiv0906.3804L}.
  The expectation of \eqref{Qtheta} w.r.t.\ $h$, conditioned on $f_t$, is from \eqref{expM2}
  \begin{eqnarray}  \nonumber  \mathbb E [\nu_{\mathcal Q}|f_t] =\int_{D} {\mathcal M}^{\alpha}_t(z)  G(z) dz
 =\int_{D_t} {\mathcal M}^{\alpha}_0(w)N_t(w) dw ,
 \end{eqnarray}
where ${\mathcal M}_0^{\alpha}(w)=|w| (\Im w)^{-\kappa/8}$ is the (unconditioned) free boundary GFF expectation $\mathbb E\, e^{\alpha {{h}}(w)}$.
Finally, taking  expectation  w.r.t.\ $f_t$ gives via the martingale $\mathcal M_t^{\alpha}(z)$  the expected quantum length in $D$ (here $\vartheta:=\arg z$):
 $$
\mathbb E \nu_{\mathcal Q} (D)=\int_D dz \mathcal M_0^{\alpha}(z) G(z)=\int_{D}(\sin \vartheta)^{8/\kappa-2} dz;
$$
for $\kappa=4$, it coincides with the \textit{Euclidean area} of $D$.

\textit{SLE boundary quantum measure.}---Consider now the reverse Schramm-Loewner map $f_t(x)$ restricted  to $x\in f_t^{-1}(\mathbb R_+\setminus K_t)$ on the real axis (i.e.,  to the right of $A$ in  Fig. \ref{slegffD}), such that $f_t'(x) \geq 1$ \cite{MR2153402}. The boundary analogs of the exponential martingales  \eqref{expM} are
\begin{eqnarray} \label{expMbound}
\hat {\mathcal M}^{\beta}_t(x):=\mathbb E\big(e^{\beta h(x)}|f_t\big)=e^{\beta
{\mathfrak h}_t(x)} [f_t'(x)]^{-\beta^2},
\end{eqnarray}
for any real $\beta$, such that $\mathbb E   \hat {\mathcal M}^{\beta}_t(x)= \hat {\mathcal M}^{\beta}_0(x)= x^{{2\beta/\sqrt{\kappa}}}$. From  \eqref{ht}  one has
$\hat {\mathcal M}_t^\beta(x)=u^{2\beta/\sqrt{\kappa}}{f'_t}(x)^{\hat d}
$
with $u:=f_t(x)$ and $\hat d=\beta Q -\beta^2$, the boundary analog of KPZ \eqref{xalpha} \cite{2008arXiv0808.1560D}.

A  \textit{boundary} fractal measure $\hat \nu$, supported on the intersection of a chordal $\mathrm{SLE}_\kappa$ curve $X=\eta$
with the axis $\mathbb R$, for $\kappa\in (4,8)$, has been constructed recently \cite{MR2430703,*2008arXiv0810.0940A}. For any interval $I \subset \mathbb R_+$, its expectation is the simple integral $\hat\nu(I)=\int_I x^{\hat d-1} dx$, where $\hat d:=2-{8}/{\kappa}$  is the Hausdorff boundary dimension of
 $X \cap \mathbb R$ \cite{MR2430703}. As in  \eqref{cov}, under the  map $f_t$,  the expected measure $\hat\nu_{f_t} (I_t )$ of the intersection of the image path $f_t({X})$ with the image interval $I_t:=f_t(I)$ is  conformally covariant:
 $\hat\nu_{f_t} (I_t)=\int_{I_t} \hat N_t(u) du$,
where $\hat N_t(u):= (xf'_t(x))^{\hat d -1}, x=f_t^{-1}(u)$.
Replacing $f_t^{-1}$ by $f_{-t}$, we observe that $\hat M_t:=  (f_{-t}/f'_{-t})^{\hat d-1}$ describes the density of expected boundary measure of $X=\eta$
\textit{given} $\eta([0,t])$, and $\hat M_t$ is a local martingale w.r.t.\ $f_{-t}$ \cite{2008arXiv0810.0940A}.

	 We define the expected SLE \textit{quantum boundary measure} $\hat\nu_{\mathcal Q}$ as
$$\hat\nu_{\mathcal Q} (I,h):=\int_I e^{\beta h(x)} x^{\hat d -1} dx ,$$
where $\beta:=\sqrt \kappa/2-2/\sqrt \kappa$  satisfies the boundary KPZ  relation above
 for  $\hat d$ (and the boundary Seiberg bound $\beta \leq Q/2$ \cite{1990PThPS.102..319S,2009arXiv0901.0277D}); its conditional expected measure
 \begin{eqnarray}  \nonumber \hat \nu_{\mathcal Q}|f_t:=\int_{I} \big(e^{\beta h(x)}|f_t\big)  x^{\hat d -1} dx\stackrel{\rm in\,\, law}{=}
  \int_{I_t} e^{\beta h(u)}\hat N_t(u) du
 \end{eqnarray}
 yields by Doob-Meyer decomposition the SLE quantum boundary measure, with now $\mathbb E [e^{\beta h(x)+\beta h(y)}\hat M_t(x) \hat M_t(y)]$  bounded from \cite{MR2430703,*2008arXiv0810.0940A}  by  $|x-y|^{\hat {\mathfrak d}-1}$, with  $\hat {\mathfrak d}:=\hat d-2\beta^2=6-\kappa/2-16/\kappa$, which is integrable for $\hat {\mathfrak d}>0$, i.e., $\kappa \in (4,8)$.
 The expectation with respect to $h$ conditioned on $f_t$  is
\begin{eqnarray}\nonumber\mathbb E [\hat\nu_{\mathcal Q} (I,h)|f_t]=\int_{I} {\hat {\mathcal M}}^{\beta}_t(x) x^{\hat d -1} dx
 =\int_{I_t} {\hat {\mathcal M}}^{\beta}_0(u)\hat N_t(u) du ,
 \end{eqnarray}
where $\hat {\mathcal M}_0^{\beta}(u)=u^{1-4/\kappa}$.
Taking expectation
 w.r.t.\ $f_t$, we find for $\kappa \in (4,8)$ that
$
\mathbb E\hat\nu_{\mathcal Q}(I)=\int_I x^{2-12/\kappa} dx;
$
for $\kappa=6$, this coincides with the \textit{Euclidean length} of $I$.

Finally, the expected \textit{quantum boundary length} $d\mathcal L:= dx\, \mathbb  E [\exp(\hat\beta h(x))|f_t]$,  is obtained for $\hat d=1$ in the above, with $\hat \beta=\gamma/2$, as expected \cite{2008arXiv0808.1560D}, and  with the deterministic forms   $d{\mathcal L}=u\, du$ for $\kappa \leq 4$, and $d{\mathcal L}= u^{4/\kappa}\, du$ for $\kappa > 4$.

We provided a foundational relationship between SLE, KPZ  and Liouville quantum gravity. We hope it will help  to solve the outstanding open problem of relating these objects to discrete models and random planar maps.

\begin{acknowledgments}
Support by grants ANR-08-BLAN-0311-CSD5, CNRS-PEPS-PTI 2010 and NSF grants DMS 0403182/064558 and OISE 0730136 is gratefully acknowledged.
\end{acknowledgments}
\bibliography{slekpz}

%merlin.mbs apsrev4-1.bst 2010-07-25 4.21a (PWD, AO, DPC) hacked
%Control: key (0)
%Control: author (72) initials jnrlst
%Control: editor formatted (1) identically to author
%Control: production of article title (-1) disabled
%Control: page (0) single
%Control: year (1) truncated
%Control: production of eprint (0) enabled
\begin{thebibliography}{34}%
\makeatletter
\providecommand \@ifxundefined [1]{%
 \@ifx{#1\undefined}
}%
\providecommand \@ifnum [1]{%
 \ifnum #1\expandafter \@firstoftwo
 \else \expandafter \@secondoftwo
 \fi
}%
\providecommand \@ifx [1]{%
 \ifx #1\expandafter \@firstoftwo
 \else \expandafter \@secondoftwo
 \fi
}%
\providecommand \natexlab [1]{#1}%
\providecommand \enquote  [1]{``#1''}%
\providecommand \bibnamefont  [1]{#1}%
\providecommand \bibfnamefont [1]{#1}%
\providecommand \citenamefont [1]{#1}%
\providecommand \href@noop [0]{\@secondoftwo}%
\providecommand \href [0]{\begingroup \@sanitize@url \@href}%
\providecommand \@href[1]{\@@startlink{#1}\@@href}%
\providecommand \@@href[1]{\endgroup#1\@@endlink}%
\providecommand \@sanitize@url [0]{\catcode `\\12\catcode `\$12\catcode
  `\&12\catcode `\#12\catcode `\^12\catcode `\_12\catcode `\%12\relax}%
\providecommand \@@startlink[1]{}%
\providecommand \@@endlink[0]{}%
\providecommand \url  [0]{\begingroup\@sanitize@url \@url }%
\providecommand \@url [1]{\endgroup\@href {#1}{\urlprefix }}%
\providecommand \urlprefix  [0]{URL }%
\providecommand \Eprint [0]{\href }%
\providecommand \doibase [0]{http://dx.doi.org/}%
\providecommand \selectlanguage [0]{\@gobble}%
\providecommand \bibinfo  [0]{\@secondoftwo}%
\providecommand \bibfield  [0]{\@secondoftwo}%
\providecommand \translation [1]{[#1]}%
\providecommand \BibitemOpen [0]{}%
\providecommand \bibitemStop [0]{}%
\providecommand \bibitemNoStop [0]{.\EOS\space}%
\providecommand \EOS [0]{\spacefactor3000\relax}%
\providecommand \BibitemShut  [1]{\csname bibitem#1\endcsname}%
\let\auto@bib@innerbib\@empty
%</preamble>
\bibitem [{\citenamefont {Polyakov}(1981)}]{MR623209}%
  \BibitemOpen
  \bibfield  {author} {\bibinfo {author} {\bibfnamefont {A.~M.}\ \bibnamefont
  {Polyakov}},\ }\href@noop {} {\bibfield  {journal} {\bibinfo  {journal}
  {Phys. Lett. B}\ }\textbf {\bibinfo {volume} {103}},\ \bibinfo {pages} {207}
  (\bibinfo {year} {1981})}\BibitemShut {NoStop}%
\bibitem [{\citenamefont {Knizhnik}\ \emph {et~al.}(1988)\citenamefont
  {Knizhnik}, \citenamefont {Polyakov},\ and\ \citenamefont
  {Zamolodchikov}}]{MR947880}%
  \BibitemOpen
  \bibfield  {author} {\bibinfo {author} {\bibfnamefont {V.~G.}\ \bibnamefont
  {Knizhnik}}, \bibinfo {author} {\bibfnamefont {A.~M.}\ \bibnamefont
  {Polyakov}}, \ and\ \bibinfo {author} {\bibfnamefont {A.~B.}\ \bibnamefont
  {Zamolodchikov}},\ }\href@noop {} {\bibfield  {journal} {\bibinfo  {journal}
  {Mod. Phys. Lett. A}\ }\textbf {\bibinfo {volume} {3}},\ \bibinfo {pages}
  {819} (\bibinfo {year} {1988})}\BibitemShut {NoStop}%
\bibitem [{\citenamefont {David}(1988)}]{MR981529}%
  \BibitemOpen
  \bibfield  {author} {\bibinfo {author} {\bibfnamefont {F.}~\bibnamefont
  {David}},\ }\href@noop {} {\bibfield  {journal} {\bibinfo  {journal} {Modern
  Phys. Lett. A}\ }\textbf {\bibinfo {volume} {3}},\ \bibinfo {pages} {1651}
  (\bibinfo {year} {1988})}\BibitemShut {NoStop}%
\bibitem [{\citenamefont {Distler}\ and\ \citenamefont
  {Kawai}(1989)}]{MR1005268}%
  \BibitemOpen
  \bibfield  {author} {\bibinfo {author} {\bibfnamefont {J.}~\bibnamefont
  {Distler}}\ and\ \bibinfo {author} {\bibfnamefont {H.}~\bibnamefont
  {Kawai}},\ }\href@noop {} {\bibfield  {journal} {\bibinfo  {journal} {Nucl.
  Phys. B}\ }\textbf {\bibinfo {volume} {321}},\ \bibinfo {pages} {509}
  (\bibinfo {year} {1989})}\BibitemShut {NoStop}%
\bibitem [{\citenamefont {{Kazakov}}(1986)}]{1986PhLA..119..140K}%
  \BibitemOpen
  \bibfield  {author} {\bibinfo {author} {\bibfnamefont {V.~A.}\ \bibnamefont
  {{Kazakov}}},\ }\href {\doibase 10.1016/0375-9601(86)90433-0} {\bibfield
  {journal} {\bibinfo  {journal} {Phys. Lett. A}\ }\textbf {\bibinfo {volume}
  {119}},\ \bibinfo {pages} {140} (\bibinfo {year} {1986})}\BibitemShut
  {NoStop}%
\bibitem [{\citenamefont {{Duplantier}}\ and\ \citenamefont
  {{Sheffield}}()}]{2008arXiv0808.1560D}%
  \BibitemOpen
  \bibfield  {author} {\bibinfo {author} {\bibfnamefont {B.}~\bibnamefont
  {{Duplantier}}}\ and\ \bibinfo {author} {\bibfnamefont {S.}~\bibnamefont
  {{Sheffield}}},\ }\href@noop {} {\bibfield  {journal} {\bibinfo  {journal}
  {Invent. math. (to appear)}\ }}\Eprint {http://arxiv.org/abs/0808.1560}
  {arXiv:0808.1560} \BibitemShut {NoStop}%
\bibitem [{\citenamefont {{Duplantier}}\ and\ \citenamefont
  {{Sheffield}}(2009)}]{2009arXiv0901.0277D}%
  \BibitemOpen
  \bibfield  {author} {\bibinfo {author} {\bibfnamefont {B.}~\bibnamefont
  {{Duplantier}}}\ and\ \bibinfo {author} {\bibfnamefont {S.}~\bibnamefont
  {{Sheffield}}},\ }\href {\doibase 10.1103/PhysRevLett.102.150603} {\bibfield
  {journal} {\bibinfo  {journal} {Phys. Rev. Lett.}\ }\textbf {\bibinfo
  {volume} {102}},\ \bibinfo {pages} {150603} (\bibinfo {year}
  {2009})}\BibitemShut {NoStop}%
\bibitem [{\citenamefont {Belavin}\ \emph {et~al.}(1984)\citenamefont
  {Belavin}, \citenamefont {Polyakov},\ and\ \citenamefont
  {Zamolodchikov}}]{MR757857}%
  \BibitemOpen
  \bibfield  {author} {\bibinfo {author} {\bibfnamefont {A.~A.}\ \bibnamefont
  {Belavin}}, \bibinfo {author} {\bibfnamefont {A.~M.}\ \bibnamefont
  {Polyakov}}, \ and\ \bibinfo {author} {\bibfnamefont {A.~B.}\ \bibnamefont
  {Zamolodchikov}},\ }\href@noop {} {\bibfield  {journal} {\bibinfo  {journal}
  {Nucl. Phys. B}\ }\textbf {\bibinfo {volume} {241}},\ \bibinfo {pages} {333}
  (\bibinfo {year} {1984})}\BibitemShut {NoStop}%
\bibitem [{\citenamefont {{Nienhuis}}(1984)}]{1984JSP....34..731N}%
  \BibitemOpen
  \bibfield  {author} {\bibinfo {author} {\bibfnamefont {B.}~\bibnamefont
  {{Nienhuis}}},\ }\href {\doibase 10.1007/BF01009437} {\bibfield  {journal}
  {\bibinfo  {journal} {J. Stat. Phys.}\ }\textbf {\bibinfo {volume} {34}},\
  \bibinfo {pages} {731} (\bibinfo {year} {1984})}\BibitemShut {NoStop}%
\bibitem [{\citenamefont {Schramm}(2000)}]{MR1776084}%
  \BibitemOpen
  \bibfield  {author} {\bibinfo {author} {\bibfnamefont {O.}~\bibnamefont
  {Schramm}},\ }\href@noop {} {\bibfield  {journal} {\bibinfo  {journal}
  {Israel J. Math.}\ }\textbf {\bibinfo {volume} {118}},\ \bibinfo {pages}
  {221} (\bibinfo {year} {2000})}\BibitemShut {NoStop}%
\bibitem [{\citenamefont {{Ginsparg}}\ and\ \citenamefont
  {{Moore}}(1993)}]{Ginsparg-Moore}%
  \BibitemOpen
  \bibfield  {author} {\bibinfo {author} {\bibfnamefont {P.}~\bibnamefont
  {{Ginsparg}}}\ and\ \bibinfo {author} {\bibfnamefont {G.}~\bibnamefont
  {{Moore}}},\ }in\ \href@noop {} {\emph {\bibinfo {booktitle} {Recent
  direction in particle theory, Proceedings of the 1992 TASI}}},\ \bibinfo
  {editor} {edited by\ \bibinfo {editor} {\bibfnamefont {J.}~\bibnamefont
  {{Harvey}}}\ and\ \bibinfo {editor} {\bibfnamefont {J.}~\bibnamefont
  {{Polchinski}}}}\ (\bibinfo  {publisher} {World Scientific},\ \bibinfo
  {address} {Singapore},\ \bibinfo {year} {1993})\BibitemShut {NoStop}%
\bibitem [{\citenamefont {{Duplantier}}(2000)}]{2000PhRvL..84.1363D}%
  \BibitemOpen
  \bibfield  {author} {\bibinfo {author} {\bibfnamefont {B.}~\bibnamefont
  {{Duplantier}}},\ }\href {\doibase 10.1103/PhysRevLett.84.1363} {\bibfield
  {journal} {\bibinfo  {journal} {Phys. Rev. Lett.}\ }\textbf {\bibinfo
  {volume} {84}},\ \bibinfo {pages} {1363} (\bibinfo {year}
  {2000})}\BibitemShut {NoStop}%
\bibitem [{\citenamefont {Duplantier}(2003)}]{MR1964687}%
  \BibitemOpen
  \bibfield  {author} {\bibinfo {author} {\bibfnamefont {B.}~\bibnamefont
  {Duplantier}},\ }\href@noop {} {\bibfield  {journal} {\bibinfo  {journal} {J.
  Stat. Phys.}\ }\textbf {\bibinfo {volume} {110}},\ \bibinfo {pages} {691}
  (\bibinfo {year} {2003})}\BibitemShut {NoStop}%
\bibitem [{\citenamefont {{Bauer}}\ and\ \citenamefont
  {{Bernard}}(2003)}]{2003CMaPh.239..493B}%
  \BibitemOpen
  \bibfield  {author} {\bibinfo {author} {\bibfnamefont {M.}~\bibnamefont
  {{Bauer}}}\ and\ \bibinfo {author} {\bibfnamefont {D.}~\bibnamefont
  {{Bernard}}},\ }\href {\doibase 10.1007/s00220-003-0881-x} {\bibfield
  {journal} {\bibinfo  {journal} {Commun. Math. Phys.}\ }\textbf {\bibinfo
  {volume} {239}},\ \bibinfo {pages} {493} (\bibinfo {year}
  {2003})}\BibitemShut {NoStop}%
\bibitem [{\citenamefont {{Friedrich}}\ and\ \citenamefont
  {{Werner}}(2003)}]{2003CMaPh.243..105F}%
  \BibitemOpen
  \bibfield  {author} {\bibinfo {author} {\bibfnamefont {R.}~\bibnamefont
  {{Friedrich}}}\ and\ \bibinfo {author} {\bibfnamefont {W.}~\bibnamefont
  {{Werner}}},\ }\href {\doibase 10.1007/s00220-003-0956-8} {\bibfield
  {journal} {\bibinfo  {journal} {Commun. Math. Phys.}\ }\textbf {\bibinfo
  {volume} {243}},\ \bibinfo {pages} {105} (\bibinfo {year}
  {2003})}\BibitemShut {NoStop}%
\bibitem [{\citenamefont {{Sheffield}}()}]{sheffield-2010}%
  \BibitemOpen
  \bibfield  {author} {\bibinfo {author} {\bibfnamefont {S.}~\bibnamefont
  {{Sheffield}}},\ }\href@noop {} {\bibinfo  {journal} {in preparation}\
  }\BibitemShut {NoStop}%
\bibitem [{\citenamefont {{Klevtsov}}()}]{2007arXiv0709.3664K}%
  \BibitemOpen
\bibfield  {journal} {  }\bibfield  {author} {\bibinfo {author} {\bibfnamefont
  {S.}~\bibnamefont {{Klevtsov}}},\ }\href@noop {} {}\Eprint
  {http://arxiv.org/abs/0709.3664} {arXiv:0709.3664} \BibitemShut {NoStop}%
\bibitem [{\citenamefont {{Bettelheim}}\ and\ \citenamefont
  {{Wiegmann}}(2009)}]{BW}%
  \BibitemOpen
  \bibfield  {author} {\bibinfo {author} {\bibfnamefont {E.}~\bibnamefont
  {{Bettelheim}}}\ and\ \bibinfo {author} {\bibfnamefont {P.}~\bibnamefont
  {{Wiegmann}}},\ }\href@noop {} {} (\bibinfo {year} {2009}),\ \bibinfo {note}
  {at \textit{Facets of Integrability} (Paris)}\BibitemShut {NoStop}%
\bibitem [{\citenamefont {{Astala}}\ \emph {et~al.}(2010)\citenamefont
  {{Astala}}, \citenamefont {{Jones}}, \citenamefont {{Kupiainen}},\ and\
  \citenamefont {{Saksman}}}]{AJKS}%
  \BibitemOpen
  \bibfield  {author} {\bibinfo {author} {\bibfnamefont {K.}~\bibnamefont
  {{Astala}}}, \bibinfo {author} {\bibfnamefont {P.}~\bibnamefont {{Jones}}},
  \bibinfo {author} {\bibfnamefont {A.}~\bibnamefont {{Kupiainen}}}, \ and\
  \bibinfo {author} {\bibfnamefont {E.}~\bibnamefont {{Saksman}}},\ }\href@noop
  {} {\bibfield  {journal} {\bibinfo  {journal} {C. R. Acad. Sci. Paris S\'er.
  I Math.}\ }\textbf {\bibinfo {volume} {348}},\ \bibinfo {pages} {257}
  (\bibinfo {year} {2010})}\BibitemShut {NoStop}%
\bibitem [{\citenamefont {Sheffield}(2007)}]{MR2322706}%
  \BibitemOpen
  \bibfield  {author} {\bibinfo {author} {\bibfnamefont {S.}~\bibnamefont
  {Sheffield}},\ }\href@noop {} {\bibfield  {journal} {\bibinfo  {journal}
  {Probab. Th. Rel. Fields}\ }\textbf {\bibinfo {volume} {139}},\ \bibinfo
  {pages} {521} (\bibinfo {year} {2007})}\BibitemShut {NoStop}%
\bibitem [{\citenamefont {Rohde}\ and\ \citenamefont
  {Schramm}(2005)}]{MR2153402}%
  \BibitemOpen
  \bibfield  {author} {\bibinfo {author} {\bibfnamefont {S.}~\bibnamefont
  {Rohde}}\ and\ \bibinfo {author} {\bibfnamefont {O.}~\bibnamefont
  {Schramm}},\ }\href@noop {} {\bibfield  {journal} {\bibinfo  {journal} {Ann.
  of Math.}\ }\textbf {\bibinfo {volume} {161}},\ \bibinfo {pages} {883}
  (\bibinfo {year} {2005})}\BibitemShut {NoStop}%
\bibitem [{\citenamefont {Lawler}\ \emph {et~al.}(2004)\citenamefont {Lawler},
  \citenamefont {Schramm},\ and\ \citenamefont {Werner}}]{MR2044671}%
  \BibitemOpen
  \bibfield  {author} {\bibinfo {author} {\bibfnamefont {G.~F.}\ \bibnamefont
  {Lawler}}, \bibinfo {author} {\bibfnamefont {O.}~\bibnamefont {Schramm}}, \
  and\ \bibinfo {author} {\bibfnamefont {W.}~\bibnamefont {Werner}},\
  }\href@noop {} {\bibfield  {journal} {\bibinfo  {journal} {Ann. Probab.}\
  }\textbf {\bibinfo {volume} {32}},\ \bibinfo {pages} {939} (\bibinfo {year}
  {2004})}\BibitemShut {NoStop}%
\bibitem [{\citenamefont {Smirnov}(2010)}]{MR2680496}%
  \BibitemOpen
  \bibfield  {author} {\bibinfo {author} {\bibfnamefont {S.}~\bibnamefont
  {Smirnov}},\ }\href {\doibase 10.4007/annals.2010.172.1441} {\bibfield
  {journal} {\bibinfo  {journal} {Ann. of Math. (2)}\ }\textbf {\bibinfo
  {volume} {172}},\ \bibinfo {pages} {1435} (\bibinfo {year}
  {2010})}\BibitemShut {NoStop}%
\bibitem [{\citenamefont {{Chelkak}}\ and\ \citenamefont
  {{Smirnov}}()}]{2009arXiv0910.2045C}%
  \BibitemOpen
  \bibfield  {author} {\bibinfo {author} {\bibfnamefont {D.}~\bibnamefont
  {{Chelkak}}}\ and\ \bibinfo {author} {\bibfnamefont {S.}~\bibnamefont
  {{Smirnov}}},\ }\href@noop {} {\bibinfo  {journal} {arXiv:0910.2045}\
  }\BibitemShut {NoStop}%
\bibitem [{\citenamefont {Schramm}\ and\ \citenamefont
  {Sheffield}(2009)}]{MR2486487}%
  \BibitemOpen
\bibfield  {journal} {  }\bibfield  {author} {\bibinfo {author} {\bibfnamefont
  {O.}~\bibnamefont {Schramm}}\ and\ \bibinfo {author} {\bibfnamefont
  {S.}~\bibnamefont {Sheffield}},\ }\href@noop {} {\bibfield  {journal}
  {\bibinfo  {journal} {Acta Math.}\ }\textbf {\bibinfo {volume} {202}},\
  \bibinfo {pages} {21} (\bibinfo {year} {2009})}\BibitemShut {NoStop}%
\bibitem [{\citenamefont {Smirnov}(2001)}]{MR1851632}%
  \BibitemOpen
  \bibfield  {author} {\bibinfo {author} {\bibfnamefont {S.}~\bibnamefont
  {Smirnov}},\ }\href@noop {} {\bibfield  {journal} {\bibinfo  {journal} {C. R.
  Acad. Sci. Paris S\'er. I Math.}\ }\textbf {\bibinfo {volume} {333}},\
  \bibinfo {pages} {239} (\bibinfo {year} {2001})}\BibitemShut {NoStop}%
\bibitem [{\citenamefont {{Klebanov}}(1995)}]{1995PhRvD..51.1836K}%
  \BibitemOpen
  \bibfield  {author} {\bibinfo {author} {\bibfnamefont {I.}~\bibnamefont
  {{Klebanov}}},\ }\href@noop {} {\bibfield  {journal} {\bibinfo  {journal}
  {Phys. Rev. D}\ }\textbf {\bibinfo {volume} {51}},\ \bibinfo {pages} {1836}
  (\bibinfo {year} {1995})}\BibitemShut {NoStop}%
\bibitem [{\citenamefont {{Lawler}}\ and\ \citenamefont
  {{Sheffield}}()}]{2009arXiv0906.3804L}%
  \BibitemOpen
  \bibfield  {author} {\bibinfo {author} {\bibfnamefont {G.~F.}\ \bibnamefont
  {{Lawler}}}\ and\ \bibinfo {author} {\bibfnamefont {S.}~\bibnamefont
  {{Sheffield}}},\ }\href@noop {} {\ }\Eprint {http://arxiv.org/abs/0906.3804}
  {arXiv:0906.3804} \BibitemShut {NoStop}%
\bibitem [{\citenamefont {Beffara}(2008)}]{MR2435854}%
  \BibitemOpen
  \bibfield  {author} {\bibinfo {author} {\bibfnamefont {V.}~\bibnamefont
  {Beffara}},\ }\href@noop {} {\bibfield  {journal} {\bibinfo  {journal} {Ann.
  Probab.}\ }\textbf {\bibinfo {volume} {36}},\ \bibinfo {pages} {1421}
  (\bibinfo {year} {2008})}\BibitemShut {NoStop}%
\bibitem [{\citenamefont {{Seiberg}}(1990)}]{1990PThPS.102..319S}%
  \BibitemOpen
  \bibfield  {author} {\bibinfo {author} {\bibfnamefont {N.}~\bibnamefont
  {{Seiberg}}},\ }\href@noop {} {\bibfield  {journal} {\bibinfo  {journal}
  {Progr. Theor. Phys. Suppl.}\ }\textbf {\bibinfo {volume} {102}},\ \bibinfo
  {pages} {319} (\bibinfo {year} {1990})}\BibitemShut {NoStop}%
\bibitem [{\citenamefont {{Lawler}}\ and\ \citenamefont
  {{Zhou}}()}]{2010arXiv1006.4936L}%
  \BibitemOpen
  \bibfield  {author} {\bibinfo {author} {\bibfnamefont {G.~F.}\ \bibnamefont
  {{Lawler}}}\ and\ \bibinfo {author} {\bibfnamefont {W.}~\bibnamefont
  {{Zhou}}},\ }\href@noop {} {\ }\Eprint {http://arxiv.org/abs/1006.4936}
  {arXiv:1006.4936} \BibitemShut {NoStop}%
\bibitem [{\citenamefont {{Lawler}}\ and\ \citenamefont {{Werness}}()}]{LWess}%
  \BibitemOpen
  \bibfield  {author} {\bibinfo {author} {\bibfnamefont {G.~F.}\ \bibnamefont
  {{Lawler}}}\ and\ \bibinfo {author} {\bibfnamefont {B.~M.}\ \bibnamefont
  {{Werness}}},\ }\href@noop {} {\ }\Eprint {http://arxiv.org/abs/1011.3551}
  {arXiv:1011.3551} \BibitemShut {NoStop}%
\bibitem [{\citenamefont {Alberts}\ and\ \citenamefont
  {Sheffield}(2008)}]{MR2430703}%
  \BibitemOpen
  \bibfield  {author} {\bibinfo {author} {\bibfnamefont {T.}~\bibnamefont
  {Alberts}}\ and\ \bibinfo {author} {\bibfnamefont {S.}~\bibnamefont
  {Sheffield}},\ }\href@noop {} {\bibfield  {journal} {\bibinfo  {journal}
  {Electron. J. Probab.}\ }\textbf {\bibinfo {volume} {13}},\ \bibinfo {pages}
  {1166} (\bibinfo {year} {2008})}\BibitemShut {NoStop}%
\bibitem [{\citenamefont {{Alberts}}\ and\ \citenamefont
  {{Sheffield}}()}]{2008arXiv0810.0940A}%
  \BibitemOpen
  \bibfield  {author} {\bibinfo {author} {\bibfnamefont {T.}~\bibnamefont
  {{Alberts}}}\ and\ \bibinfo {author} {\bibfnamefont {S.}~\bibnamefont
  {{Sheffield}}},\ }\href@noop {} {\ }\Eprint {http://arxiv.org/abs/0810.0940}
  {arXiv:0810.0940} \BibitemShut {NoStop}%
\end{thebibliography}%
%\bibliography{slekpz}
\bibliographystyle{apsrev4-1}
\end{document}